\documentclass[letterpaper, 10 pt, conference]{ieeeconf}  
                              \pdfoutput=1                            

\IEEEoverridecommandlockouts                              
\usepackage{fancyhdr}
\newcommand{\mytitle}{\textbf{Accepted final version.}
To appear in \textit{Proc. of the 57th IEEE International Conference on Decision and Control (CDC), 2018}.\\
\copyright 2018 IEEE. Personal use of this material is permitted. Permission
from IEEE must be obtained for all other uses, in any current or future
media, including reprinting/republishing this material for advertising or
promotional purposes, creating new collective works, for resale or
redistribution to servers or lists, or reuse of any copyrighted component of
this work in other works.}
\usepackage{cite}
\usepackage{balance}
\usepackage{xspace} 
\usepackage{url}
\usepackage[caption=false,font=footnotesize]{subfig}
\usepackage{paralist}
\usepackage{fixltx2e}
\usepackage[pdftex]{graphicx}  
\usepackage{amsmath,amssymb}
\usepackage{xfrac}
\usepackage{array}
\usepackage{algorithm,algorithmic}
\usepackage{textcomp} 
\usepackage{siunitx}
\usepackage{pgfplots}
\usepackage{xcolor}
\pgfplotsset{compat=newest,unit code/.code={\si{#1}},plot coordinates/math parser=false,grid style={lightgray}}
\usepgfplotslibrary{units,external,groupplots}
\usetikzlibrary{positioning,angles,quotes,patterns}

\newcommand{\norm}[1]{\left\lVert#1\right\rVert}
\DeclareMathOperator*{\E}{\mathbb{E}}

\newcommand\figref[1]{Fig.~\ref{#1}}

\newcommand\secref[1]{Sec.~\ref{#1}}

\newcommand{\eg}{\emph{e.g.},\xspace}
\newcommand{\ie}{\emph{i.e.},\xspace}

\usepackage{ifthen} 
\newboolean{authnotes} 

\setboolean{authnotes}{true} 

\ifthenelse{\boolean{authnotes}}
{
\newcommand{\jj}[1]{\footnote{{\bf\color{blue} JJ: #1}}}
\newcommand{\db}[1]{\footnote{{\bf\color{blue} Dominik: #1}}} 
\newcommand{\st}[1]{\footnote{{\bf\color{green!50!black} Sebastian: #1}}}
\newcommand{\gm}[1]{\footnote{{\bf\color{blue} Georg: #1}}}
}
{
\newcommand{\jj}[1]{}
\newcommand{\db}[1]{}
\newcommand{\st}[1]{}
\newcommand{\gm}[1]{}
}
\newcommand{\rulesep}{\unskip\ \vrule\ }

\title{\LARGE \bf 
Deep Reinforcement Learning for Event-Triggered Control
} 

\author{Dominik Baumann$^{1,*}$, Jia-Jie Zhu$^{2,*}$, Georg Martius$^{2}$, and Sebastian Trimpe$^{1}$
\thanks{$^{*}$Equal contribution.}
\thanks{$^{1}$Intelligent Control Systems Group, Max Planck Institute for Intelligent Systems, Stuttgart/T\"{u}bingen, Germany. 
Email: \{dominik.baumann, sebastian.trimpe\}@tuebingen.mpg.de.}%
\thanks{$^{2}$Autonomous Learning Group, Max Planck Institute for Intelligent Systems, T\"{u}bingen, Germany.
Email:  \{jia-jie.zhu, georg.martius\}@tuebingen.mpg.de.}%
\thanks{This work was supported in part by the German Research Foundation (DFG) within the priority program SPP 1914 (grant TR 1433/1-1), 
        the European Union’s Horizon 2020 research and innovation programme under the Marie Sklodowska-Curie grant agreement No 798321,
         the Cyber Valley Initiative, and the Max Planck Society.}
}

\begin{document}

\maketitle 
\thispagestyle{fancy}	
\pagestyle{empty}

\begin{abstract}

Event-triggered control (ETC) methods can achieve high-performance control with a significantly lower number of samples compared to usual, time-triggered methods. 
These frameworks are often based on a mathematical model of the system and specific designs of controller and event trigger. 
In this paper, we show how deep reinforcement learning (DRL)  algorithms can be leveraged to simultaneously learn control and communication behavior from scratch, and present a DRL approach that is particularly suitable for ETC. 
To our knowledge, this is the first work to apply DRL to ETC. 
We validate the approach on multiple control tasks and compare it to model-based event-triggering frameworks. 
In particular, we demonstrate that it can, other than many model-based ETC designs, be straightforwardly applied to nonlinear systems.

\end{abstract}

\section{Introduction}
\label{sec:intro}

In modern engineering systems, feedback loops are often closed over (wired or wireless) communication networks~\cite{hespanha2007survey,lunze2014control}. 
Examples for these networked control systems (NCSs) include smart buildings, where sensors and actuators are deployed to regulate the indoor climate; swarms of drones, which exchange information for coordinated flight; 
or autonomous driving, where communication between the vehicles allows for adaptive traffic control.
When multiple control loops use the same network as in these examples, communication becomes a shared and therefore limited resource.
Classical control methods typically ignore this fact and take sufficient communication resources for granted.  Data is exchanged in a time-triggered fashion between components of the control loops irrespective of whether an update is actually needed.
In recent years, the research community in event-triggered control (ETC) has had remarkable success in showing that the amount of samples in feedback loops can be reduced significantly compared to those time-triggered approaches (see experimental studies \cite{TrDAn11,araujo2014system,dolk2017event} for example).
In ETC, transmission of data and thus closing of the feedback loop is triggered only on certain events, \eg an error growing too large.  This way, feedback happens only when necessary, and significant communication savings can be realized.

There exists a large variety of methods to design event-triggered controllers, see \eg~\cite{HeJoTa,Mi15} for an overview.
Most of these methods are rooted in classical control theory and based on a model of the process to be controlled.
In contrast to this, we show herein that \emph{deep reinforcement learning} (DRL) algorithms can be leveraged in order to learn both control \emph{and} communication law from scratch without the need for a dynamics model.
We formulate resource-aware control as a reinforcement learning (RL) problem, where the learning agent optimizes its actions (control input and communication decision) so as to maximize some expected reward over a time horizon.  The reward function is composed of two terms, one capturing control performance, and one that gives rewards for time steps without communication.
This way, the agent learns to control the system with good performance, but without communicating all the time.

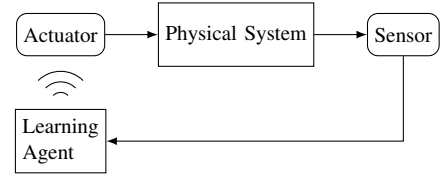
\begin{figure}
\centering
\begin{tikzpicture}[remember picture,scale=0.8, every node/.style={scale=0.8}]
\tikzset{>=latex}
\tikzset{radiation/.style={{decorate,decoration={expanding
waves,angle=90,segment length=4pt}}}}
\node[draw,rectangle,rounded corners,minimum width=2em, minimum
height=2em](act_et){Actuator};
\node[draw, rectangle,minimum width = 3em, minimum height =
3em,right = 2em of act_et](system_et){Physical System};
\node[draw,rectangle,rounded corners,minimum width=2em, minimum
height=2em,right = 2em of system_et](sensor_et){Sensor};
\node[draw,rectangle, minimum width = 3em, minimum height = 3em, below = 2em of
act_et,align = left](ctrl_et){Learning \\ Agent};

\draw[->](system_et) -- (sensor_et);
\draw[->](sensor_et) |- (ctrl_et);
\draw[->] (act_et) -- (system_et);
\draw[radiation,decoration={angle=45}]
([shift={(0cm,0.1cm)}]ctrl_et.north)-- 
([shift={(0cm,-0.1cm)}]act_et.south);
\end{tikzpicture}
\caption{Learning of event-triggered control. The learning agent continuously receives sensor inputs, but has to transmit control signals over a resource-limited wireless network. The agent learns both control and communication; that is, (i) what actuator command to send, and (ii) when to send it. 
}
\label{fig:RL_oneAgent}
\end{figure}

More specifically, we consider the setup  in \figref{fig:RL_oneAgent}.
While the learning agent is directly connected to the sensor and thus receives its measurements continuously, it must transmit actuator commands over a wireless network link.  Thus, we seek to reduce communication of control inputs.
We propose two approaches for learning ETC in this setting.
In the first approach, we assume that a time-triggered feedback controller is given, and we learn only the communication policy.
As an alternative, we learn both control and communication policy simultaneously.  
This can be regarded as \emph{end-to-end learning}.  
In DRL, end-to-end learning (\eg \cite{levine2016end}) typically refers to learning the complete control policy from raw sensor data to actuator commands `end to end,' without (artificially) separating into sub-tasks such as filtering, planning, and tracking.    
In the context of ETC, end-to-end thus emphasizes learning of both control and communication simultaneously, rather than separating the two.
This is particularly interesting as the separation principle does not generally hold in ETC; 
that is, optimizing controller and communication structure separately, as often done in practice, does not necessarily yield the overall optimal event-triggered control law~\cite{RaSaBaJo}. End-to-end DRL is a way to overcome this separation.




By means of numerical examples, we demonstrate that end-to-end learning of ETC is feasible.  Moreover, we compare to some common model-based ETC approaches. 
The comparison reveals that, for linear settings with an accurate model available, model-based ETC typically cannot be outperformed by the proposed DRL approach---at least, at medium to high average communication rates.  In some cases, however, DRL can find superior policies at very low communication rates, where model-based ETC yields unstable solutions.  
In contrast to common ETC methods, the proposed learning approach straightforwardly applies also to nonlinear control problems.

\paragraph*{Contributions}
The contributions of this work can be summarized as follows:
\begin{itemize}
\item Proposal of deep reinforcement learning (DRL) to learn event-triggered controllers from data;
\item learning of communication policy only (with a given controller) with policy gradients~\cite{peters2008reinforcement};
\item end-to-end learning of control \emph{and} communication policy with deep deterministic policy gradient (DDPG) algorithm~\cite{lillicrap2015continuous};
\item demonstration of feasibility of DRL in numerical benchmark problems; and 
\item comparison to model-based ETC methods.
\end{itemize}

\paragraph*{Related work} 
Using machine learning techniques to learn feedback controllers from data has been considered in previous works, see \eg \cite{kober2013reinforcement,schaal2010learning,MaHeScTr17,doerr_ICRA_2017,baumann2018memristor,abbeel2010autonomous,levine2016end,calandra2016bayesian,khargonekar2018advancing,lillicrap2015continuous,peters2008reinforcement} and references therein.
These works typically consider learning of control policies  only, without incorporating the cost of communication such as when controller and plant are connected over a network link.

There exists a large body of work on ETC methods, see \eg~\cite{HeJoTa,Mi15} and references therein for an overview. 
Using RL for ETC is not discussed there and has generally received less attention.
Model-free RL for event-triggered controllers has for example been proposed in~\cite{VaFe}, where an actor-critic method is used to learn an event-triggered controller and stability of the resulting system is proved.
However, the authors consider a predefined communication trigger (a threshold on the difference between current and last communicated state); that is, they do not learn the communication policy from scratch.
Similarly, in~\cite{ZhNiHeXu}, an approximate dynamic programming approach using neural networks is implemented to learn event-triggered controllers, again with a fixed error threshold for triggering communication.
In~\cite{sahoo2016neural}, the authors propose an algorithm to update the weights of a neural network in an event-triggered fashion.
Model-based RL is used in~\cite{YaHeLi17} to simultaneously learn an optimal event-triggered controller with a predefined fixed communication threshold, and a model of the system.  To the authors' knowledge, no prior work considers end-to-end learning of control and communication in ETC, which is the main contribution herein.


In~\cite{so18}, learning is proposed to improve communication behavior for event-triggered state estimation.
Other than here, the idea is to improve accuracy of state predictions through model-learning.
A second event-trigger is introduced that triggers learning experiments only if the mathematical model deviates from the real system.

\paragraph*{Outline}
The next section continues with a short introduction to ETC and a more detailed description of DRL with particular focus on approaches for continuous state-action spaces.
The proposed approaches for DRL of ETC are then introduced in \secref{sec:approach}.  Section \ref{sec:validation} presents numerical results, and the paper concludes with a discussion in \secref{sec:conclusion}.

\section{Background}
\label{sec:backgr}
We consider a nonlinear, discrete-time system disturbed by additive Gaussian noise,
\begin{subequations}
\label{eqn:sys}
\begin{align}
x_{k+1} &= f\left(x_k,u_k\right)+v_k\\
y_k &=x_k+w_k,
\end{align}
\end{subequations}
with $k$ the discrete-time index, $x_k,y_k,v_k,w_k\in\mathbb{R}^{n}$, $u_k\in\mathbb{R}^l$, and $v_k,w_k$  mutually independent Gaussian random variables with probability density functions (PDFs) $\mathcal{N}\left(v_k;0,\Sigma_\mathrm{p}\right)$ and $\mathcal{N}\left(w_k;0,\Sigma_\mathrm{m}\right)$, and variances $\Sigma_\mathrm{p}$ and $\Sigma_\mathrm{m}$.

\subsection{Event-triggered Control}
\label{sec:ETC}
In ETC communication is not triggered by a clock, but by the occurrence of certain events.
The input is defined as
\begin{align}
\label{eqn:input}
u_k = \begin{cases}
\mathcal{K}_k\left(x_k\right)&\text{ if }\gamma_k = 1\\
u_{k-1} &\text{ if }\gamma_k = 0
\end{cases}
\end{align}
where $\gamma_k$ denotes the communication decision and $\mathcal{K}_k$ the control law.
Whether to communicate is decided based on a triggering law $\mathcal{C}_k$,
\begin{align}
\label{eqn:EBC_update}
\gamma_k = 1 \iff \mathcal{C}_k\left(x_k,\hat{x}_k\right)\ge 0,
\end{align}
where $\hat{x}_k$ defines the state of the system at the last time instant a control input has been applied.
An example for such a triggering law would be
\begin{align}
\label{eqn:EBC_trigExample}
\mathcal{C}_k\left(x_k,\hat{x}_k\right)\ge 0 \iff \norm{x_k-\hat{x}_k}_2\ge \delta,
\end{align}
with $\delta$ being a predefined threshold.
Intuitively speaking this would mean we communicate the control input if the current state of the system deviates too much from the state at the last communication slot.
There are many other ETC schemes following similar ideas; we refer to~\cite{HeJoTa,Mi15} for more detailed overviews.

In this work, we want to learn both the control law $\mathcal{K}_k$ and the triggering policy ($\gamma_k$ as a function of the observables) using DRL approaches.

\subsection{Deep Reinforcement Learning}
\label{sec:DRL}
We give a brief introduction to RL in general and present the two baseline algorithms we later focus on in \secref{sec:approach}.

The main goal in RL is to learn an optimal policy by trial and error while interacting with the environment.
Mathematically, this can be formulated as a Markov decision process (MDP).
In an MDP, we consider the setting where an agent interacts with the environment. 
At every time step, the agent selects an action $a_k$, from the action space $A$, based on its current state $s_k$, from the state space $S$, according to a policy $\pi\left(a_k\vert s_k\right)$.\footnote{If we want to learn a controller for a dynamical system, often $s_k\equiv x_k$ and $a_k\equiv u_k$ holds. However, this is not necessarily the case and, in particular, not in the setup we shall develop herein.} 
The agent receives a reward $r_k$ and the state transitions to the next state $s_{k+1}$ according to the state transition probability $p(s',r\vert s\!=\!s_k,a\!=\!a_k)$.
The goal of the RL agent is to maximize the expected discounted reward $ \mathbb{E} [R_k] = \mathbb{E}\left[\sum_{i=0}^{\mathrm{T}-1} \zeta^ir_{k+i}\right]$, where $\zeta\in\left(0,1\right]$ is the discount factor.\footnote{To simplify the formulation, we consider the episodic case with $k\in[0,T-1]$.}
There are generally two types of RL methods: model-free and model-based.
One model-free method to achieve the goal is to learn a value function $v_\pi(s) \overset{_!}{=} \mathbb{E}\left[R_k\vert s_k = s\right]$, which denotes the expected return in case policy $\pi$ is followed from state $s$ onwards.
The value function $v_\pi(s)$ follows the Bellman equation~\cite{bellman1957dynamic},
\begin{align}
\label{eqn:bellman}
v_\pi(s) = \sum\limits_a \pi\left(a\vert s\right)\sum\limits_{s',r} p\left(s',r\vert s,a\right)\left[r+\zeta v_\pi(s')\right],
\end{align}
which can then be maximized to find the optimal state values.

Similarly, one can estimate a state-action value function (Q-function) $Q_\pi\left(s,a\right)\overset{_!}{=} \E\left(R_k\vert s_k=s,a_k=a\right)$ which determines the expected return for selecting action $a$ in state $s$ and following the policy $\pi$ thereafter.
In an MDP, the optimal action in the current state can be derived by maximizing the Q-function.
If the transition probabilities $p$ are available, this can, \eg be done using (exact) dynamic programming (DP).
In cases where the model is not known, we resort to RL (simulation-based approximate DP) methods.
In such cases, the Q-function can be learned using the Q-learning algorithm presented in~\cite{watkins1992q},
\begin{align}
\label{eqn:q-learning}
\begin{split}
Q(s_k,a_k) &\leftarrow Q(s_k,a_k)
\\&+\alpha\left(r_k+\zeta\max_{a_k'}Q\left(a_k',s_k'\right)-Q(s_k,a_k)\right).
\end{split}
\end{align}
The Q-learning algorithm updates the Q-function using the collected experience $(s_k,a_k,r_k,s_k')$.
For a more detailed introduction to RL, see~\cite{sutton1998reinforcement}.

This basic RL approach has successfully been applied to low-dimensional tasks with discrete state and action space~\cite{sutton1998reinforcement}.
For controlling a dynamical system, we usually deal with a continuous state and action space, which might be of high dimension for complex systems.
Continuous spaces could be discretized, but the discretization needs to be very fine for high-performance control.
This, in turn, leads to very high-dimensional state and action spaces imposing unreasonable computational complexity and hampering convergence speed drastically.

To the rescue come parametrized function approximators.
In machine learning, deep neural networks (DNNs) have widely been used to handle high-dimensional tasks.
Recently, they have also been applied to RL, giving rise to the field of DRL.
For instance, in deep Q-learning~\cite{mnih2015human}, the state-action function $Q$ is approximated with a DNN, making it possible to solve complex tasks in high-dimensional continuous state spaces.
However, this algorithm only works for discrete action spaces.

One possible solution to this problem
is the actor-critic architecture~\cite{sutton1998reinforcement}.
The actor outputs continuous actions while the critic estimates the value function.
Both can be implemented using DNNs.
One such algorithm is deep deterministic policy gradient (DDPG)~\cite{lillicrap2015continuous,silver2014deterministic}, which we introduce in the following.

As an alternative, we look at policy search methods that directly learn a policy without a Q-function.
Specifically, we will present the policy gradient algorithm~\cite{peters2008reinforcement} and the trust region policy optimization (TRPO) algorithm~\cite{schulman2015:trpo}.

\subsubsection{DDPG}
The DDPG algorithm, and a variation of it, are presented in~\cite{lillicrap2015continuous,hausknecht2015deep}.
For completeness, we restate the main derivations.

DDPG is an actor-critic algorithm with two networks.
One is the actor network $\mu$, parametrized by $\theta^\mu$ that takes the state $s_k$ as input and outputs an action $a_k$.
Additionally, we have the critic network $Q$, parametrized by $\theta^Q$, which takes state and action as input and outputs a scalar estimate of the value function, the Q-value $Q\left(s_k,a_k\right)$.
The updates of the critic network are close to the original formulation of the Q-learning algorithm given in~\eqref{eqn:q-learning}.
Adapting~\eqref{eqn:q-learning} to the described neural network setting leads to minimizing the loss function
\begin{align}
\label{eqn:loss_critic}
\begin{split}
L_Q\left(s_k,a_k\vert \theta^Q\right) &= \left(\vphantom{\max_{a_k'}} Q\left(s_k,a_k\vert\theta^Q\right) \right. \\
&\left. -\left(r_k+\zeta\max_{a_k'}Q\left(s_k',a_k'\vert\theta^Q\right) \right) \right)^2 .
\end{split}
\end{align}
For continuous action spaces, equation~\eqref{eqn:loss_critic} is not tractable, as we would have to maximize over the next-state action $a_k'$.
Instead, we take the next-state action $a_k' = \mu\left(s_k'\vert \theta^\mu\right)$ of the actor network.
Inserting this in equation~\eqref{eqn:loss_critic} leads to
\begin{align}
\label{eqn:loss_actorcritic}
\begin{split}
L_Q\left(s_k,a_k\vert \theta^Q\right) &= \left(Q\left(s_k,a_k\vert\theta^Q\right) \right. \\
&\left. -\left(r_k+\zeta Q\left(s_k',\mu\left(s_k'\vert \theta^\mu\right)\vert\theta^Q\right) \right) \right)^2 .
\end{split}
\end{align}
Based on this loss function, the critic can learn the value function via gradient descent.
Clearly, a crucial point is the quality of the actor's policy.
The actor tries to minimize the difference between its current output $a$ and the optimal policy $a^*$,
\begin{align}
\label{eqn:actor}
L_\mu\left(s_k\vert\theta^\mu\right)=\left(a_k-a_k^*\right)=\left(\mu\left(s_k\vert \theta^Q\right)-a_k^* \right)^2.
\end{align}
The true optimal action $a_k^*$ is of course unknown. 
As simply estimating it would require to solve a global optimization problem in continuous space, the critic network can instead provide a gradient that leads to higher estimated Q-values: $\nabla_{a_k} Q\left(s_k,a_k\vert \theta^Q\right)$.
Computing this gradient is much faster. 
This was first introduced in~\cite{Hafner2011}.
The gradient implies a change in actions, which is used to update the actor network in this direction by backpropagation.
In particular, for an observed state $s_k$ and action $a_k$, the parameters of the actor network are changed according to
\begin{align}
\label{eqn:actor_update}
\nabla_{\theta^\mu}J = \nabla_{a_k}Q\left(s_k,a_k\vert\theta^Q\right)\nabla_{\theta^\mu}\mu\left(s_k\vert\theta^\mu\right)
\end{align}
 approximating the minimization of~\eqref{eqn:actor}.

Two general problems arise from this approach.
For most optimization algorithms, it is usually assumed that samples are independent and identically distributed.
This is obviously not the case if we sequentially explore an environment.
To resolve this, a replay buffer of fixed size that stores tuples $\left(s_k,a_k,r_k,s_{k+1}\right)$ is used.
Actor and critic are now updated by uniformly sampling mini-batches from this replay buffer.

 The second problem is that the update of the Q-network uses the current Q-network to compute the target values (see~\eqref{eqn:loss_actorcritic}).
This has proved to be unstable in many environments.
Therefore, copies of actor and critic networks, $Q'\left(s_k,a_k\vert\theta^{Q'}\right)$ and $\mu'\left(s_k\vert\theta^{\mu'}\right)$ are created and used to calculate the target values.
The copies are updated by slowly tracking the learned network,
\begin{align}
\label{eqn:learn_copy}
\theta' = \kappa\theta + \left(1-\kappa\right)\theta',
\end{align}
with $\kappa\ll 1$.
This typically leads to more robust learning.

In \secref{sec:ParamAct}, we will show how this algorithm can be used to jointly learn controller and communication behavior.

\subsubsection{Trust Region Policy Optimization (TRPO)}
\label{sec:PG}
A second approach is to perform direct parameter search without a value function. This is referred to as direct policy search or policy gradient~\cite{peters2008reinforcement}.
A parametrized policy $\pi$ is adapted directly to maximize the expected reward.
Since, without a model, analytical gradients of the reward function are not available, policy gradient methods use stochastic policies and adapt them to increase the likelihood of a high reward.
Formally, let the policy $\pi(s_k;\theta^\pi)$ representing $p(a_k|s_k)$ be parametrized by $\theta^\pi$ in a differentiable way.
Now we aim to maximize the utility $J(\theta^\pi) = \mathbb{E}_{a_k\sim \pi(s_k;\theta^\pi)}R(s_k,a_k)$.
Policy gradient methods follow the gradient estimator of $J$ for a given trajectory:
\begin{align}
\label{eqn:PG}
\nabla_{\theta^\pi} J(\theta^\pi) &= \sum_{k=0}^{T-1} R_k \nabla_{\theta^\pi} \log \pi(s_k;\theta^\pi)\,.
\end{align}

A recent advance of policy gradient methods is given by the Trust Region Policy Optimization~\cite{schulman2015:trpo} (TRPO) that uses a surrogate optimization objective and a trust region approach for updating the policy efficiently. 
In terms of theoretical guarantees, this algorithm ensures monotonic improvement of the policy performance, given the amount of training samples is large.
We will use this method in \secref{sec:Gate} to learn the controller and triggering policy independently.


\section{Approach}
\label{sec:approach}
We present two approaches to learn resource-aware control.
First, we consider learning communication structure and controller end-to-end.
The policy should then output both, the communication decision and the control input,
\begin{align} 
\label{eqn:DDPG_policy}
\left(\gamma_k,u_k\right) = \pi_{\mathrm{combined}}\left(s_k\right) = \pi_\mathrm{combined}\left(x_k,u_{k-1}\right),
\end{align}
where $\gamma_k$ is a binary variable with $\gamma_k=0$ indicating no communication.
Alternatively, we start with a control strategy for the system without communication constraints, either learned or designed. 
The goal is then to learn the communication structure, \ie a policy 
\begin{align}
\label{eqn:gate_policy}
\gamma_k=\pi_{\mathrm{comm}}\left(s_k\right) = \pi_\mathrm{comm}\left(x_k,u_k,u_{k-1}\right).
\end{align}
This strategy requires us to separate the design of controller and communication structure. 

For both settings, the state of the RL agent includes the current state $x_k$ of the system and the last control input $u_{k-1}$.
This is necessary, as in case of no communication, $u_{k-1}$ will be applied again, so knowledge of the last control input is needed for the problem to form an MDP.
In~\eqref{eqn:gate_policy}, the state is further augmented and also includes the current control input $u_k$.
The RL agent learns a communication policy, \ie it needs to decide, whether $u_k$ or $u_{k-1}$ will be applied.
Therefore it needs knowledge of both.
The action $a_k$ of the RL agent consists of the communication decision for the separated policy, and of communication decision and control input in the combined case.

In RL, the reward function typically depends on the states and actions of the system.
We additionally consider communication, thus we arrive at a reward function of the form
\begin{align}
\label{eqn:reward}
r_k = -x_k^\top Qx_k-u_k^\top Ru_k-\lambda\gamma_k,
\end{align}
where $\lambda$ is a hyper-parameter.
During training, the agent receives negative rewards for bad performance and for communication.
In an episodic reinforcement learning task, where agents' interaction with the environment is divided into episodes, an additional constant positive reward is often given to the agent to prevent undesired early termination of the episodes, \eg the pole dropping for the cart-pole system.

\subsection{Joint Learning of Communication and Control}
\label{sec:ParamAct}
To learn resource-aware controllers, we consider both the discrete action space (the decision whether to communicate) and the continuous action space (the control input that should be applied).
This is related to the idea of reinforcement learning in parameterized action space \cite{masson2016reinforcement, hausknecht2015deep}.

This framework considers a parameterized action space Markov decision process (PAMDP), which involves a set of discrete actions $A_\mathrm{d}=\left\{d_1,d_2,\ldots,d_k\right\}$.
Each discrete action $d\in A_\mathrm{d}$ is associated with $m_\mathrm{d}$ continuous parameters $\left\{p_1^\mathrm{d},p_2^\mathrm{d},\ldots,p_{m_\mathrm{d}}^\mathrm{d}\right\}\in\mathbb{R}^{m_\mathrm{d}}$. 
An action is represented by a tuple $\left(d,p_1^\mathrm{d},\ldots,p_{m_\mathrm{d}}^\mathrm{d}\right)$.
This leads to the following action space: $A = \cup_{\mathrm{d}\in\mathrm{A_\mathrm{d}}} \left(d,p_1^\mathrm{d},\ldots,p_{m_\mathrm{d}}^\mathrm{d}\right)$.

In our case, there are two discrete actions, $d_1$ and $d_2$, where $d_1$ corresponds to the decision to communicate the control input ($\gamma_k = 1$).
Accordingly, the action $d_1$ has $m_\mathrm{d_1}=1$ continuous parameter $p_1^\mathrm{d_1}=u_k$, which is the control input.
Action $d_2$ does not have any continuous parameter, as we apply the last control input again.

\begin{figure}
\centering
\includegraphics[scale=0.5]{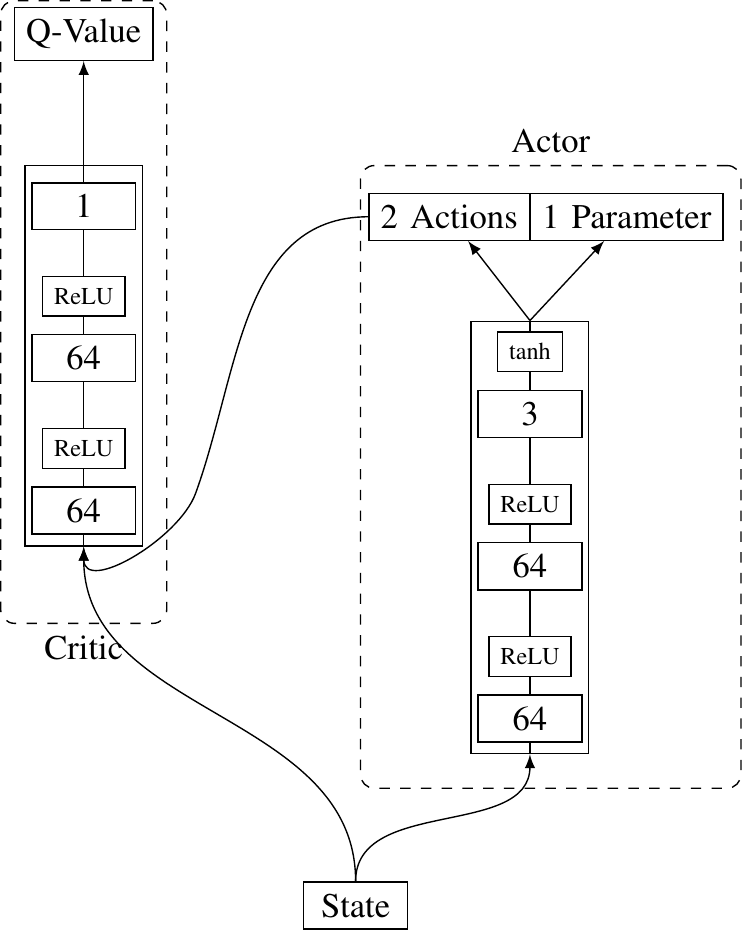}
\rulesep 
\includegraphics[scale=0.5]{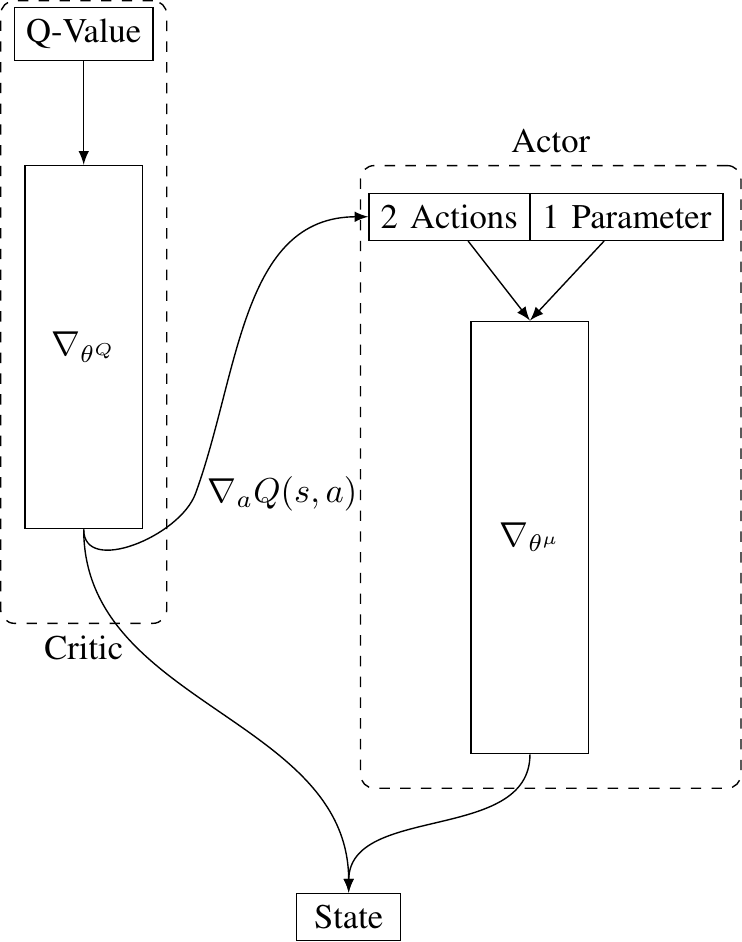} 
\vspace{-0.2cm}
\caption{Visualization of the actor-critic network structure adapted from~\cite{hausknecht2015deep}. 
On the left, the general network architecture, showing the units and activation function of each layer.
Each block represents one layer of the network with the number describing the number of neurons.
The smaller blocks indicate the activation functions. 
On the right, the update of the actor using back-propagation.}
\label{fig:network}
\end{figure}

As stated in \secref{sec:DRL}, we consider the DDPG algorithm\footnote{Our implementation is based on the non-parameterized DDPG framework provided in~\cite{baselines}.}, where both actor and critic network are implemented using DNNs. 
The architecture is depicted in \figref{fig:network}.
The actor network outputs continuous values for all actions in the action space, \ie we have
\begin{align}
\label{eqn:param_policy}
a_k = \left(d_1,d_2,u_k\right) = \pi_\mathrm{combined}\left(x_k,u_{k-1}\right).
\end{align}
This is different from~\eqref{eqn:DDPG_policy}, as we do not receive a discrete parameter $\gamma_k$, but continuous values for all parameters $a_k$.
To obtain a discrete decision, we determine the communication decision by
\begin{align}
\gamma_k = \begin{cases}
1 &\text{ if }d_{1,k} > d_{2,k}\\
0 &\text{ otherwise}.
\end{cases}
\label{decision}
\end{align}
The continuous parameter (the control input $u_k$) is directly obtained as an output of the actor.
The output of the actor and the current state then serve as input for the critic, which estimates the Q-function value.
This structure has been applied to a gaming environment in~\cite{hausknecht2015deep}.


During training, exploration is done in an $\epsilon$-greedy fashion.
With probability $\epsilon$, we select a random discrete action (whether to communicate).
Besides the $\epsilon$-greedy exploration we add exploration noise in form of an Ornstein Uhlenbeck process to the output of the actor as has been successfully demonstrated in~\cite{lillicrap2015continuous}. 
Pseudo-code of this approach is presented in Algorithm~\ref{alg:joint}.

\begin{algorithm}
\small
\caption{Jointly learn communication and controller (adapted from~\cite{lillicrap2015continuous}).}
\label{alg:joint}
\begin{algorithmic}[1]
\STATE Initialize $\epsilon$
\STATE Randomly initialize critic DNN $Q\left(s_k,a_k\vert \theta^Q\right)$ and actor DNN $\mu\left(s_k\vert\theta^\mu\right)$ with weights $\theta^Q$ and $\theta^\mu$.
\STATE Initialize target networks $Q'$ and $\mu'$ with weights $\theta^{Q'}\leftarrow\theta^Q$, $\theta^{\mu'}\leftarrow\theta^\mu$
\STATE Initialize replay buffer $\mathcal{R}$
\FOR {episode $=1$ to $N$}
\STATE Receive initial observation state $s_1$
\FOR {k $=1$ to $M$}
\STATE Generate uniformly distributed $\phi\in\left[0,1\right]$ 
\IF {$\phi<\epsilon$}
\STATE Generate $\xi\sim B\left(2,0.5\right)$ from Bernoulli distribution
\IF {$\xi==1$}
\STATE Choose discrete action $d_1$
\ELSE
\STATE Choose discrete action $d_2$
\ENDIF
\ELSE
\STATE Select $a_k = \mu\left(s_k\vert\theta^Q\right)$ and apply exploration noise to the actor output
\STATE Get communication decision $\gamma_k$ using~\eqref{decision}
\ENDIF
\STATE Execute action $a_k$, receive reward $r_k$ and state $s_{k+1}$
\STATE Store transition $\left(s_k,a_k,r_k,s_{k+1}\right)$ in $\mathcal{R}$
\STATE Sample random mini-batch from $\mathcal{R}$
\STATE Update critic by minimizing loss function~\eqref{eqn:loss_actorcritic}
\STATE Update actor policy using sampled policy gradient~\eqref{eqn:actor_update}
\STATE Update target networks according to equation~\eqref{eqn:learn_copy}
\ENDFOR
\ENDFOR
\end{algorithmic}
\end{algorithm}
\subsection{Learning Communication only}
\label{sec:Gate}
An alternative to the aforementioned end-to-end approach is to separately learn the communication strategy and the stabilizing controller.
In this approach, a control policy is first fully trained using a high-performing RL algorithm, \eg TRPO~\cite{schulman2015:trpo}.
Instead of hand-engineering the communication strategy, we propose to use policy gradients~\cite{peters2008reinforcement} to learn this communication structure.
In essence, the trained controller computes the control input in every time-step, whereas another learning agent controls whether to send this control input to the system, thus implementing~\eqref{eqn:gate_policy}.

The general scheme is related to hierarchical reinforcement learning~\cite{sutton1995td} and gated recurrent neural networks~\cite{chung2014empirical}.
We discuss preliminary experimental results of this alternative approach, in relatively challenging tasks, in \secref{sec:locomotion}.

\section{Validation}
\label{sec:validation}
In this section, we validate the proposed DRL approaches through several numerical simulations.
For the algorithm introduced in \secref{sec:ParamAct}, which jointly learns communication behavior and controller, we show the general applicability as a proof of concept on the inverted pendulum, compare to several model-based ETC algorithms on the same platform, and show its general applicability for nonlinear tasks.
In \secref{sec:locomotion}, we demonstrate learning resource-aware locomotion tasks using the algorithm presented in \secref{sec:Gate}.\footnote{Code of representative examples and video of resource-aware locomotion are available at \url{https://sites.google.com/view/drlcom/}.}

The numerical simulations presented in this section were carried out in environments adapted from the OpenAI Gym\footnote{\url{https://gym.openai.com/}}.
The OpenAI Gym provides simulation models of different classical control tasks, such as the inverted pendulum and the cart-pole system, as well as physics simulation systems, Atari games, and many more.
For our approaches, we augment the reward functions provided in the OpenAI Gym according to~\eqref{eqn:reward}. 
The simulations are carried out on a cluster utilizing parallel runs for the training and testing processes with randomized seeds.
\subsection{Proof of Concept}
\label{sec:poc}
As a proof of concept, we apply the learning algorithm presented in \secref{sec:ParamAct} to the inverted pendulum.
The inverted pendulum consists of a pendulum attached to a motor with the goal to keep the pendulum close to its upright position at $\theta\!=\!\SI{0}{\radian}$.
We assume process and measurement noise as in~\eqref{eqn:sys} and the initial state also a Gaussian distributed random variable with $x(0)\sim \mathcal{N}\left(x(0);0,\Sigma_0\right)$.
The standard deviation of noise and initial position was chosen to be $\num{e-4}$.

The simulation environment provides upper and lower bounds of $\pm\SI{2}{\newton\meter}$ on the input torque that may be applied to the pendulum.
One discrete time step lasts \SI{50}{\milli\second}.

We train the controller using the joint learning approach detailed in \secref{sec:ParamAct}. 
The hyper-parameter $\lambda$ in~\eqref{eqn:reward} is tuned by a grid search of 25 values between $0.01$ and $100$. Different hyper-parameter values correspond to different communication rates and controller performances.
For each hyper-parameter setting and task, we carry out 5 randomized training processes using different random seeds, each consists of one million training iterations. 
During performance evaluation, we carry out 100 randomized test episodes for each of the 5 trained agents for each hyper-parameter setting with each episode lasting 500 discrete time steps.

\begin{figure} 
 \centering
 \input{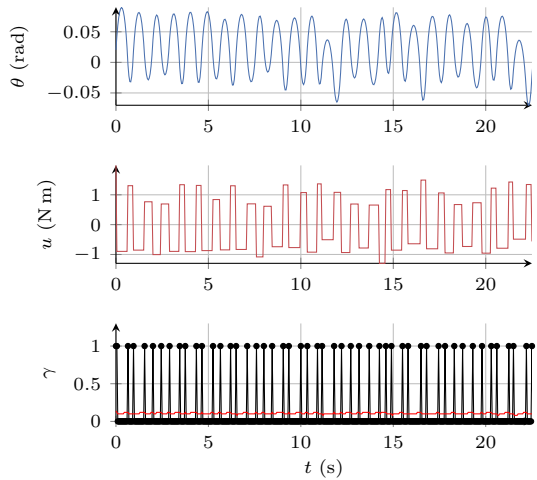}
 \vspace{-0.3cm}
 \caption{Stabilization of the inverted pendulum with an event-triggered controller learned with the method presented in \secref{sec:ParamAct}. 
 The plots show, from top to bottom, the angle of the pendulum $\theta$, the control input $u$, and the communication (decision $\gamma$ in black, average communication in red). The average communication here and in following plots is computed as a moving average over \num{50} samples.}
 \label{fig:poc}
 \end{figure}

Results of one such test episode can be seen in \figref{fig:poc}. 
The plot is a representative example for the results obtained from the different agents and test episodes.
The pendulum system remains stable with the angle staying well within $\pm\SI{0.1}{\radian}$, while significantly saving communication.
Here we observe a saving rate of around \SI{90}{\percent}.
Further it can be seen that the learning approach does not converge to a triggering law with fixed threshold.
The threshold at which communication of a new control input is triggered is dynamically changing throughout the experiment.
 
In general, stability is very important in control tasks.
However, most works in reinforcement learning aim at achieving optimal control instead of stability.
For policy iteration methods (which we use herein), guarantees on monotonic improvement of the policy can be given, as discussed for instance in~\cite{schulman2015:trpo,kakade2002approximately}.
However, the resulting controller derived in this work is approximated by a deep neural network, which is highly nonlinear, thus analyzing the stability of the system is not straightforward.
Further, finding optimal control policies does not necessarily imply stability, but the connection is more subtle (cf. \cite{kalman1960contributions,bertsekas2018stable}).
For the time being, this renders the effort of analyzing stability intractable. 
While stability of DRL is an important topic for research, this example is a proof of concept that joint control and communication policies can be found with DRL.
 
\subsection{Comparison}
\label{sec:benchmark}

We compare the performance of the learning approach to common model-based ETC designs on the inverted pendulum.
For balancing, the inverted pendulum can be approximated as a linear system and methods from linear control theory may be used.
We consider the intuitive ETC algorithm, where we only communicate, if the state deviates too much from its desired position.
The state of the inverted pendulum consists of its angle $\theta$ and its angular velocity $\dot{\theta}$, the desired value for both is zero.
Hence, we apply the following control law
\begin{align} 
\label{eqn:ETC_base}
u_k = \begin{cases}
Kx_k&\text{ if }\norm{x_k}_2>\delta\\
u_{k-1}&\text{ otherwise},
\end{cases}
\end{align}
where the matrix $K$ is designed with an LQR approach using $Q$ and $R$ as in the reward function of the learning algorithm.
Additionally, we compare to the approaches introduced in~\cite{donkers2012output} and~\cite{tabuada2007event}.
In both cases, we use the formulation as a periodic event-triggered control algorithm provided in~\cite{heemels2013periodic}, which is $\gamma_k=1 \iff \norm{K\hat{x}_k-Kx_k}>\delta\norm{Kx_k}$ for~\cite{donkers2012output}, and $\gamma_k = 1 \iff \norm{\hat{x}_k-x_k}>\delta\norm{x_k}$ for~\cite{tabuada2007event}.
The algorithms only give a communication threshold, but require a stabilizing controller $K$.
For both, we used the same LQR as for~\eqref{eqn:ETC_base}.

The algorithms have different triggering laws but are all based on a fixed threshold $\delta$.
For comparison, we vary this threshold.
After every experiment, we compute the quadratic cost and the average communication.
These simulations revealed that communication savings up to around \SI{60}{\percent} for~\cite{tabuada2007event}, \SI{70}{\percent} for~\cite{heemels2013periodic}, and \SI{80}{\percent} for~\eqref{eqn:ETC_base} are possible.
When running similar simulations with the DDPG approach from \secref{sec:ParamAct}, we noted that the model-based approaches clearly outperform the learning approach.
However, communication savings of \SI{90}{\percent}, as observed in \figref{fig:poc}, cannot be achieved with these model-based approaches as they become unstable before.
The learning agent, in contrast, is still able to come up with a good policy.





 
\subsection{Swing-up}
\label{sec:NL}
As previously stated, the presented DRL approach can also be applied to more challenging, \eg nonlinear, systems.
In this section, we take on such a setting where the aforementioned ETC designs do not apply.
The training and evaluation procedures in this section follow the same paradigm detailed in Section~\ref{sec:poc}.

\begin{figure}
\centering
\input{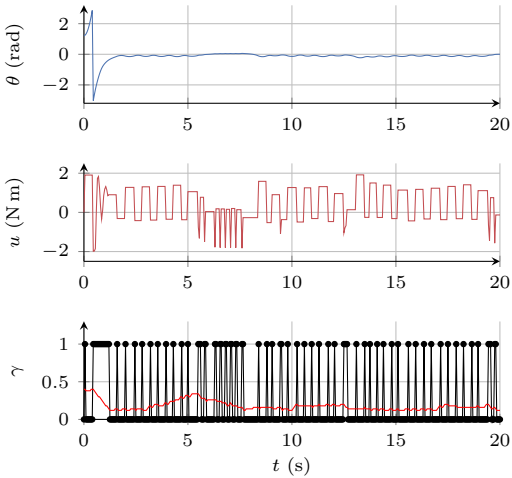}
\vspace{-0.3cm}
\caption{Resource-aware swing-up of the inverted pendulum, showing the angle $\theta$ (top) and the communication (bottom, discrete decision in black, average communication in red). 
The jump observed in the beginning is due to the pendulum crossing $\pi$ and thus immediately switching from $\pi$ to $-\pi$.}
\label{fig:swingup}
\end{figure}

In \figref{fig:swingup}, the inverted pendulum is presented again, but with the initial angle well beyond the linear region.
As can be seen, the agent is able to learn a resource-aware swing-up policy and then stabilize the pendulum around $\theta\!=\!\SI{0}{\radian}$ while saving around \SI{80}{\percent} communication.

We also trained the learning agent on the cart-pole system, where it was similarly able to learn a stable policy while saving around \SI{90}{\percent} of communication (with an underlying sample time of \SI{25}{\milli\second}).

%

\subsection{Simulated locomotion}
\label{sec:locomotion}
So far, we have addressed canonical tasks in optimal control using the proposed end-to-end approach. 
In this section, we move the focus to advanced tasks, \ie locomotion.
We applied the proposed parameterized DDPG approach of \secref{sec:ParamAct} to resource-aware locomotion, but only with moderate success. This is possibly due to the lack of reward hand-engineering and is left for future work considerations.
During our experiments, we discovered that learning controller and communication behavior separately, as explained in \secref{sec:Gate}, allows us to address even challenging tasks such as robotic locomotion.
In this approach, we first train the agent with full communication using TRPO, typically using iteration numbers on the $10^6$ order-of-magnitude. After the TRPO agent is trained, we train the communication strategy using a policy gradient approach with augmented reward as in \eqref{eqn:reward} until we observe desired behaviors trading off performance and communication saving.
Our experimental environment is based on the Mujoco physics simulation engine\cite{todorov2012mujoco}.

\begin{figure}
\centering
\includegraphics[width = 0.125\textwidth]{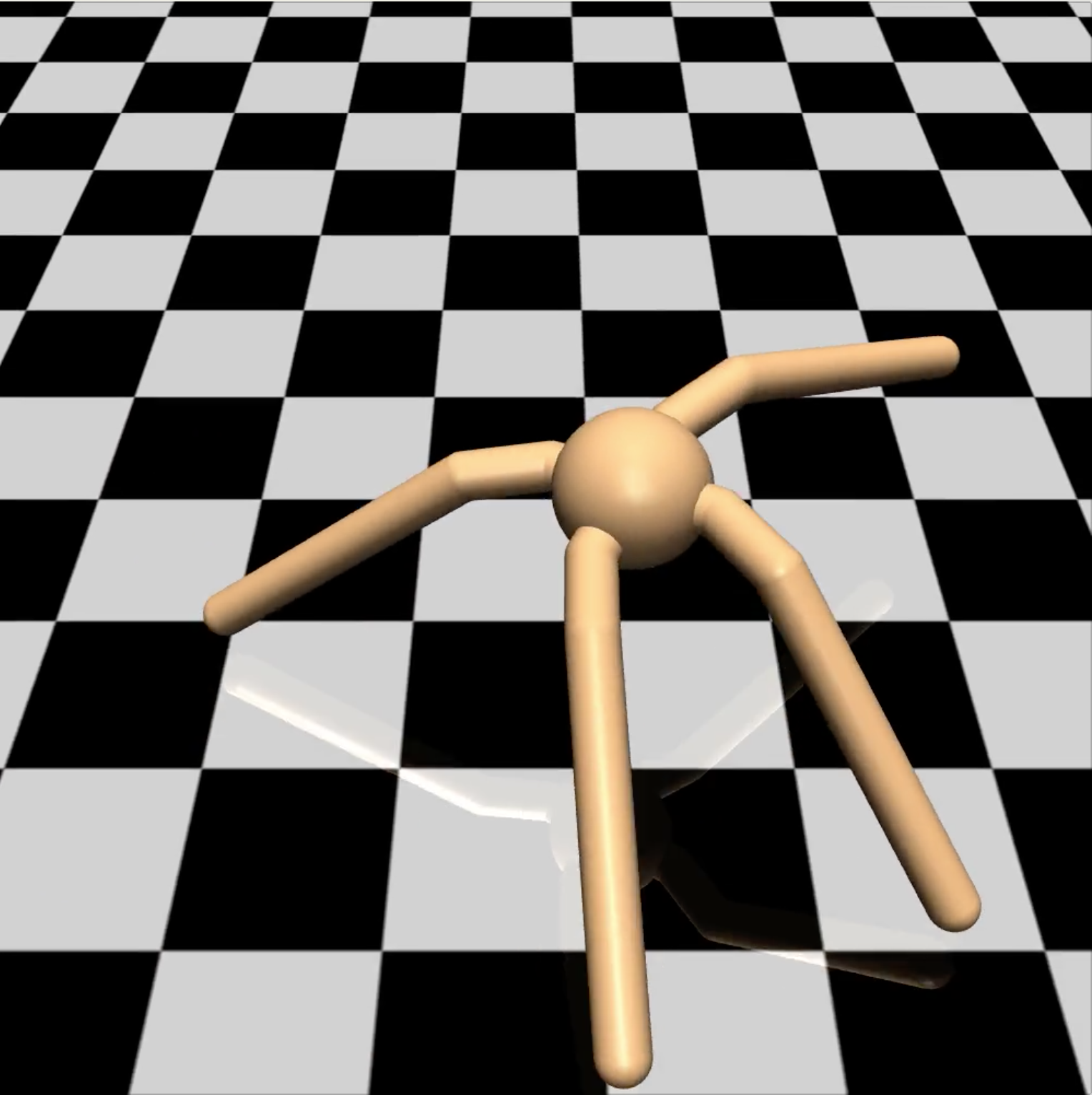}
\caption{The ant robot.}
\label{fig:Ant}
\end{figure}
As an example we trained an ant (quadruped) robot (\figref{fig:Ant}) in a simulated 3D locomotion task.
It is a relatively challenging task considering the high-dimensional state space (111 states with 13 for position, 14 for velocity, 84 for external force) and under-actuation (8 actuators). 
To make matters worse, it can be easily toppled and is then subsequently not able to stand up.
The underlying sampling time is \SI{50}{\milli\second}. 

\begin{figure}
\centering
\input{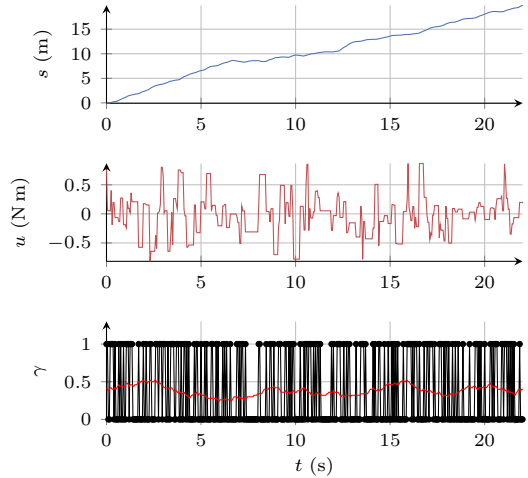}
\vspace{-0.3cm}
\caption{Simulation of the Ant robot (see \figref{fig:Ant}) learning to walk while saving communication, showing from top to bottom the position $s$ of the center of mass, the input $u$, which is the torque applied to the hip motor,  communication instants in black and the average communication in red.}
\label{fig:ant}
\end{figure}

As shown in \figref{fig:ant}, the ant learns to walk saving around \SI{60}{\percent} of communication. 
We did observe that, during some of the runs, the resource-aware ant falls and causes worse performance. 
However, this happens only around \SI{10}{\percent} of the time. 
As our method is task agnostic and not specifically engineered for locomotion tasks, we consider the performances and communication savings non-trivial.


\section{Discussion}
\label{sec:conclusion}
Most existing approaches in event-triggered control (ETC) rely on the availability of accurate dynamics models for the design of control law and event trigger.  
In contrast, we proposed the use of deep reinforcement learning (DRL) for simultaneously learning control \emph{and} communication policies from simulation data without the need of an analytical dynamics model.  
For scenarios where an accurate linear model is available, the numerical comparisons herein have shown that common model-based ETC approaches are superior to the learning approach.  
This is to be expected because the model-based design fully exploits the model structure.  
For some cases, however, the DRL approach succeeded in finding stabilizing controllers at very low average communication rates, which the model-based design were unable to obtain. 
What is more, the key advantage of the learning-based approach lies in its versatility and generality.  
As the examples herein have shown, the same algorithm can be used to also learn control and communication policies for nonlinear problems, including complex ones like locomotion.  
In the presented example, significant communication savings of around \SI{60}{\percent} were obtained.


One limitation of our current approaches is the zero-order hold (ZOH) employed at the actuator.
Instead of ZOH, some model-based approaches perform predictions based on the dynamics model in case of no communication, and thus achieve better performance.
This could also be done if learning agents are used and would lead to a two agent problem.
The first agent continuously receives measurement updates and decides when to transmit data to the second agent. 
The second agent can continuously apply control inputs, which includes the possibility of making predictions based on a learned model.
Investigating such more general learning architectures is an interesting and challenging topic for future work.
Whether theoretical guarantees such as on stability and robustness can also be obtained for the learned controllers is another topic worthwhile to be investigated.




\bibliographystyle{IEEEtran}
\bibliography{IEEEabrv,refs}

\end{document}